\documentclass{ws-ijmpcs}

\begin{document}

\markboth{J.A.~Oller}
{New insights in $\bar{K}N$ scattering and the $\Lambda(1405)$}

%
\catchline{}{}{}{}{}
%

\title{New insights in $\bar{K}N$ scattering and the $\Lambda(1405)$ 
}

\author{J.~A.~Oller}

\address{Departamento de F\'{\i}sica, Universidad de Murcia\\
Murcia, E-30071 Murcia, Spain\\
oller@um.es}

\maketitle

\begin{history}
\received{Day Month Year}
\revised{Day Month Year}
\end{history}

\begin{abstract}
We discuss in this talk aspects of $\bar{K}N$ plus coupled channels dynamics with special 
emphasis on the $\Lambda(1405)$ resonance and isovector resonances in the same 
energy region. We comment on several experimental reactions giving rise to 
$\pi \Sigma$ distributions with important implications on the spectroscopy. 
We also discuss recent theoretical models based con NLO SU(3) unitarized chiral perturbation  theory. 

\end{abstract}

\ccode{PACS numbers: 11.25.Hf, 123.1K}

\section{Introduction and interest}	

A good theoretical point in the study of S-wave $\bar{K}N$ scattering with coupled channels (we consider altogether 
ten coupled channels, $\pi^0\Lambda$, $\pi^0\Sigma^0$, $\pi^-\Sigma^+$, $\pi^+\Sigma^-$, $K^- p$, 
$\bar{K}^0 n$, $\eta\Lambda$, $\eta\Sigma^0$, $K^0 \Xi^+$ and $K^+\Xi^-$) is that the pseudoscalars involved 
are (pseudo)Goldstone bosons associated with the spontaneous chiral symmetry breaking of QCD. As a result 
we can apply chiral perturbation theory (ChPT) with baryons. However, the large masses 
associated with baryons as well as with kaons and etas, the latter are large 
  compared with the typical modulus of the $\bar{K}N$   
center of mass (c.m.) three-momentum at around the nominal mass of the $\Lambda(1405)$,
 which is around 130~MeV, give rise to non-perturbative dynamics. 
 The situation is similar to nucleon-nucleon scattering with the Deuteron or $K\bar{K}$ scattering with the 
$f_0(980)$. 

\begin{figure}[pb]
\centerline{\psfig{file=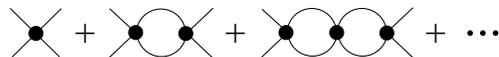,width=6.7cm}}
\vspace*{8pt}
\caption{A schematic figure illustrating the resummation of the unitarity cut to take care of the 
infrared enhancement associated with the presence of particles in the scattering with large masses. \label{chorizo}}
\end{figure}

The origin of this non-perturbative dynamics with the sizes of masses can be explained following a similar argument 
to that given by Weinberg in relation with nucleon-nucleon scattering and the nucleon mass \cite{wein2}. The point is 
that the masses of the intermediate states are large compared with the typical three-momentum involved so that 
the differences between the external energy and that of the intermediate states, that appear in the 
propagators of the intermediate states, correspond to differences between 
kinetic energies. In this way, its inverse is enhanced by an infrared factor $\sim 2M_K/q$, with $q$ the typical size of an external c.m. three-momentum and $M_K$ the kaon mass. This factor is around one order of magnitude for $|q|\sim 100$~MeV, quite close to the subthreshold three-momentum for the $\bar{K}N$ around the $\Lambda(1405)$. 
 It is also worth indicating that 
for $u$-crossed channel dynamics instead of having the difference between the external and intermediate energies one has the sum of them, so that the crossed channel loops are indeed infrared suppressed.
This implies that there is an infrared enhancement of the unitarity cut that makes 
definitely smaller the overall scale $\Lambda_{\text{ChPT}}$ over which the chiral expansion  is performed for 
meson-baryon scattering with strangeness $-1$. Then we have to resum the unitarity cut, giving rise to partial 
wave amplitudes fulfilling exactly two-body unitarity, 
while  keeping the analytical requirement associated with that cut. 
In a diagrammatic language it implies that the string of diagrams represented in Fig.~\ref{chorizo} should be resummed. 
This resummation goes beyond of perturbation theory and gives rise to the appearance of many resonances in 
meson-baryon scattering, in particular of the $\Lambda(1405)$.  In Fig.~\ref{chorizo} the filled circle represents an arbitrary meson-baryon vertex, which will be calculated perturbatively in ChPT.

The scattering of $\bar{K}N$ plus coupled channels is essential for several important problems in nuclear  physics. 
 The interested reader is referred to the recent review \cite{jidorev} for an exhaustive list of topics and 
references.

\section{Two-Pole nature for the $\Lambda(1405)$}

The first indication about the fact that the $\Lambda(1405)$ corresponds to two poles in the 
complex plane was given in Ref.~\cite{landau} by studying the pole content of the 
low-energy $\bar{K}N$ scattering amplitude in the cloudy bag model.
 The presence of two relevant poles in the energy region of the $\Lambda(1405)$ was obtained 
independently in Ref.~\cite{plb} within unitarized ChPT. This point was analyzed in 
detail later in Ref.~\cite{colla} making use of the same theoretical approach. 
Within the J\"ulich model  the two poles for the $\Lambda(1405)$ were also obtained 
in \cite{haiden}. In Ref.~\cite{plb} the pole positions for these two isoscalar poles in the second Riemann sheet (with the sign reversed 
only for the $\pi \Sigma$ three-momentum) were predicted with the values
\begin{align}
m_{low}-i\frac{1}{2}\Gamma_{low}&=1379-i\,27~\text{MeV}~,\nonumber\\
m_{high}-i\frac{1}{2}\Gamma_{high}&=1434-i\,11~\text{MeV}~.
\label{poles}
\end{align}
Notice that both poles appear between the $ \pi\Sigma$ and $\bar{K}N$ thresholds, which is precisely the physical
 energy region that connects continuously with the second Riemann sheet, where these poles lie. This is why they can 
be directly observed in experiments. Indeed, both poles have just been simultaneously 
observed in an explicitly manner, for the first time 
in any reaction,  by the CLAS Collaboration 
in  photoproduction data \cite{clas1}, with properties remarkably 
close to the numbers given above. These experimental data were not possible 
to be described at all by just one simple Breit-Wigner pole \cite{clas1}.

\begin{figure}[pb]
\centerline{\psfig{file=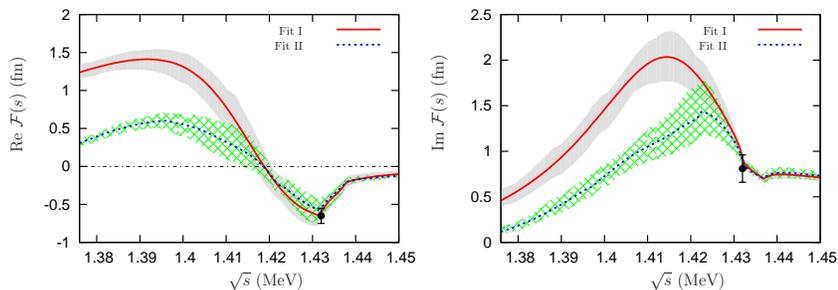,width=11.0cm}}
\vspace*{8pt}
\caption{Extrapolation of the amplitude $K^- p\to K^-p$ to the subthreshold energy region. 
\label{extrakn} }
\end{figure}

\begin{table}[ph]
\tbl{Intervals for the $I=0$ pole positions}
{\begin{tabular}{@{}lll@{}}
 \toprule
& Lower Pole [MeV] & Higher pole [MeV] \\ \colrule
Ref.~\cite{ikeda2012} & $1381^{+18}_{-6}-i\,81^{+19}_{-8}$ & $1424^{+7}_{-23}-i\,26^{+3}_{-14}$ \\
Ref.~\cite{guo2013}   & $(1380\sim 1450)-i(90\sim 150)$ & $(1413\sim 1424)-i(14\sim 31)$ \\
 \botrule
\end{tabular} \label{tablepole}}
\end{table}

The lower pole is wider because it couples more strongly with the $\pi\Sigma$ channel, while the 
higher one couples more strongly with the $\bar{K}N$ channel, whose threshold lies above in energy. 
The pole position of the lower pole is less well known that the pole position of the higher one. 
 The recent studies \cite{ikeda2012} and  \cite{guo2013}, that make use of ${\cal O}(p^2)$ ChPT 
to calculate the interacting vertices in Fig.~\ref{chorizo}, fit the available $K^-p$ two-body scattering 
data and include the most recent data on 
the energy shift and width of kaonic hydrogen \cite{sida}, find the 
intervals of pole positions shown in Table~\ref{tablepole}.
 The uncertainty in the position of the wider pole has an important impact in the subthreshold 
extrapolation of the $\bar{K}N$ scattering amplitude, a crucial input to know the properties of 
kaons in the nuclear medium \cite{bloom}.  This can be clearly seen in Fig.~\ref{extrakn}, from Ref.~\cite{guo2013},  
by the difference between the shadowed (Fit I of Ref.~\cite{guo2013}) and hatched areas
 (Fit II of Ref.~\cite{guo2013}), with the left (right) panel corresponding to the real (imaginary) part of the $K^-p\to K^-p$ S-wave scattering amplitude. 
 These two different results originate from changes  that are higher orders in the interacting vertices, calculated 
at ${\cal O}(p^2)$ in ChPT, so that they can be qualified as systematic uncertainties. It is then necessary 
to sharpen the determination of the pole position for the lower and wider pole. One option is to measure the 
$\pi\Sigma$ scattering length making use of a cusp effect in the decay $\Lambda_c\to \pi\pi\Sigma$ \cite{oka}, 
similarly as done in $\pi\pi$ from kaon decays. 
Of course, the recent results of Ref.~\cite{clas1} on photoproduction 
are also very interesting in this respect because the position of the lower pole is established around 1.37~GeV, albeit the statistics is still low.

Another strategy is to make use of the $\pi\Sigma$ event distributions and cross sections from presently available 
experimental data and reproduce all of them in a satisfactory way, once the $\bar{K}N$ cross sections in 
two-body scattering are reproduced. This has been done along the years by E.~Oset and 
collaborators. They predicted in Ref.~\cite{nacher} the event distributions for the photoproduction reaction 
$\gamma p\to K^+\pi^+\Sigma^-$ and $K^+\pi^- \Sigma^+$, which was later confirmed at LEPS \cite{ahn}. Photoproduction 
data with large statistics were recently obtained by the CLAS Collaboration \cite{moriya} and its isoscalar part 
was studied in Ref.\cite{rocaoset}. Other data correspond to the reaction $K^-p\to \pi^0\pi^0\Sigma^0$ 
 \cite{prakhov} and studied in Refs.~\cite{magas,oller2006,guo2013}. 
One also has  data from $pp$ collisions \cite{cosy,hades} 
 studied theoretically in Ref.~\cite{geng}. Old data taken in the seventies
 were also studied in Refs.~\cite{jido1,hyodo1} corresponding to $K^-d\to n \pi \Sigma$ \cite{braun} and 
$\pi^-p\to K^0 \pi \Sigma$ \cite{thomas1}, respectively. There are projects to perform similar measurements 
as the former by J-PARC E31 and IKON/KLOE DA$\Phi$NE \cite{curceanu}. It would be very interesting that in the 
future these theoretical studies could be performed by employing scattering amplitudes determined by calculating 
the interacting vertices from NLO ChPT.


\section{$I=1$ Resonances around the $\bar{K}N$ threshold}

An interesting prediction that came from the study of Ref.~\cite{plb} with unitarized ChPT 
was the existence of poles with $I=1$ around the $\bar{K}N$ threshold. This prediction 
was also confirmed by the recent study of Ref.~\cite{guo2013}, calculating the interacting vertices
 from NLO ChPT, where $I=1$ poles at $1376-i\,33$ and $1414-i\,12$ MeV were found in the second Riemann sheet.
 The $I=0$ signal is stronger than that with $I=1$, and this is the reason why these poles had not 
been observed experimentally before, as stressed in Ref.~\cite{prls}. However, 
 the $I=1$ signal  can be observed through interference 
effects with the isoscalar one. For that one needs to measure simultaneously the two charged channels $\pi^+\Sigma^-$ and 
$\pi^-\Sigma^+$.
Indeed, just by a straightforward isospin decomposition and 
neglecting the $I=2$ contribution 
(which is not resonant in these energies and  much smaller), one has
\begin{align}
\frac{d\sigma(\pi^+\Sigma^-)}{dM_I}&\propto \frac{1}{3}|T^{(0)}|^2+\frac{1}{2}|T^{(1)}|^2
+\sqrt{\frac{2}{3}}\text{Re}(T^{(0)}{T^{(1)}}^*)~,\nonumber\\
\frac{d\sigma(\pi^-\Sigma^+)}{dM_I}&\propto \frac{1}{3}|T^{(0)}|^2+\frac{1}{2}|T^{(1)}|^2
-\sqrt{\frac{2}{3}}\text{Re}(T^{(0)}{T^{(1)}}^*)~,
\end{align}
with $M_I$ the invariant mass of the $\pi\Sigma$ state. The isospin for each amplitude is indicated by the 
superscript. Notice the different sign for the interference term in $\pi^+\Sigma^-$ and $\pi^-\Sigma^+$, so that
 if one has constructive interference for one of them a destructive one results for the other. 
 It has just been recently possible to detect  with large statistics this phenomenon 
  in the photoproduction data 
from the reaction $\gamma p\to K^+\pi^+\Sigma^-$, $K^+\pi^-\Sigma^+$ and $K^+\pi^0\Sigma^0$ \cite{moriya}. As one can see in Fig.~18 of Ref.~\cite{moriya} 
there is a large shift in the peak positions of the differential cross sections 
for $\pi^+\Sigma^-$ and $\pi^-\Sigma^+$, of 
around 30~MeV, which is clearly  visible due to the large enough statistics. In addition, the height of the peak 
 also changes by around a factor of 2.
   A thorough theoretical analysis of CLAS data on $K^+\pi\Sigma$ photoproduction is necessary in order to extract with better precision the pole properties of the $J^P=1/2^-$ strangeness=$-1$ isovector resonances, first predicted in Ref.~\cite{plb}.

It is worth pointing out that the Born diagrams play an important role in the raise of these isovector resonances within 
unitarized ChPT, contrary to the $I=0$ partners that are mostly sensitive to the Weinberg-Tomozawa term \cite{ramos}.
 There is 
extra strength in the $I=1$ Born amplitudes because an SU(3) breaking mechanisms that makes 
the $u$-crossed exchanges from the $I=1$ amplitudes $\eta\Sigma\to \eta \Sigma$ and $\eta\Sigma\to K \Xi$ 
 to have their branch cut points very close the $\bar{K}N$ threshold. We also mention  that, as shown 
in Ref.~\cite{colla}, the lower pole  of the $\Lambda(1405)$ is mainly a singlet while the higher one  is an octet. 
 In addition, the $I=1$ resonances are needed to complete the two octets of $J^P=1/2^-$ strangeness=$-1$ 
resonances that are degenerated in the $SU(3)$ limit \cite{colla}, corresponding to the symmetric and antisymmetric 
octet representations of $SU(3)$. For the $I=1/2$ resonances belonging to these octets see Ref.~\cite{nieves}.

\section{The resonances around the $\bar{K}N$ threshold are dynamically generated}

Here we give the following arguments to justify that these $I=0$ and 1 resonances between the $\pi\Sigma$ and 
$\bar{K}N$ thresholds are generated by the hadron-dynamics of the meson-baryon coupled channels involved in 
the scattering. 

{\bf 1.-} The unitarity loops in the series depicted in Fig.~\ref{chorizo} require regularization, which can be 
performed by taking a once-subtracted dispersion relation with an unknown subtraction constant. 
One can then invoke a natural value for this subtraction constant as a function 
of $\Lambda_{\text{ChPT}}\simeq 1$~GeV, as deduced in 
Ref.~\cite{plb}. Denoting by $a$ the subtraction constant, its natural value is given by 
$a\simeq -2 \log\left(1+\sqrt{1+M_P^2/\Lambda^2_{\text{ChPT}}}\right)\simeq -2$, with $M_P$ a pseudoscalar mass.
 It turns out that the resonances in Ref.~\cite{plb} stem from such natural value for $a$, which corresponds 
to the one appropriate for intermediate meson-baryon states with three-momentum $\lesssim \Lambda_{\text{ChPT}}$. 

{\bf 2.-} The resonance poles corresponding to these resonances have a strong sensitivity on the actual 
Riemann sheet in which one is searching them \cite{plb}. 
Indeed, they could perfectly disappear when passing from one Riemann sheet to another. This is in  contradiction 
with the fact that resonances with a dominant bare component have pole positions which are the same 
independently of the Riemann sheet considered, as discussed in more detail in Ref.~\cite{morgan}.

{\bf 3.-} It is also interesting to calculate matter or electromagnetic radii. This was done in 
Ref.~\cite{sekihara} for the $\Lambda(1405)$ and absolute values considerably larger than those of 
the neutron were obtained, indicating a spread resonance.

{\bf 4.-} The movement of the two $I=0$ poles making up the $\Lambda(1405)$
as a function of the numbers of colors of QCD, $N_C$,  was studied in Ref.~\cite{jidoroca}. 
The resulting pole trajectories were compared with the expectation for a standard $qqq$ baryon, 
for which its mass runs as ${\cal O}(N_C)$ and its width stays fixed as ${\cal O}(N_C^0)$ \cite{goity}.  
The trajectories for the two poles of the $\Lambda(1405)$ were very different to the standard scenario,  
 indicating a different nature for these two poles.



\begin{thebibliography}{0}    

\bibitem{wein2}S.~Weinberg, Nucl. Phys. {\bf 383} (1991) 3.

\bibitem{jidorev}T.~Hyodo and D.~Jido,
  Prog.\ Part.\ Nucl.\ Phys.\  {\bf 67} (2012) 55.


\bibitem{landau}P.~J.~Frink {\it et al.}, Phys. Rev. C {\bf 41} (1990) 2720; 
 E.~A.~Veit {\it et al.}, Phys. Lett. B {\bf 137} (1984) 415.

\bibitem{plb} J.~A.~Oller and U.-G.~Mei{\ss}ner, Phys. Lett. B {\bf 500} (2001) 263.

\bibitem{colla} D.~Jido {\it et al.}, Nucl. Phys. A {\bf 725} (2003) 181.

\bibitem{haiden}J. Haidenbauer {\it et al.}, Eur. Phys. J. A {\bf 47} (2011) 18;  A.~Mueller-Groeling, K.~Holinde and J.~Speth, Nucl. Phys. A {\bf 513} (1990) 557.

\bibitem{clas1}H.~Y.~Lu {\it et al.} [CLAS Collaboration],  arXiv:1307.4411 [nucl-ex].

\bibitem{ikeda2012}Y.~Ikeda, T.~Hyodo and W.~Weise,
  Nucl.\ Phys.\ A {\bf 881} (2012) 98.

\bibitem{guo2013}Z.~-H.~Guo and J.~A.~Oller,
  Phys.\ Rev.\ C {\bf 87} (2013) 035202.

\bibitem{sida}M.~Bazzi {\it et al.} [Siddharta Collaboration], Phys. Lett. B {\bf 704} (2011) 113. 

\bibitem{bloom}S.~D.~Bloom, M.~H.~Johnson and E.~Teller, Phys. Rev. Lett. {\bf 23} (1969) 28.

\bibitem{oka}T.~Hyodo and M.~Oka, Phys. Rev. C {\bf 84} (2011) 035201.

\bibitem{nacher} C.~Nacher {\it et al.}, Phys. Lett. B {\bf 455} (1999) 55.

\bibitem{ahn}J.~K.~Ahn, Nucl. Phys. A {\bf 721} (2003) 715; M.~Niiyama {\it et al.}, Phys. Rev. C {\bf 78} (2008) 035202.

\bibitem{moriya} K.~Moriya ~{\it et al.}, Phys. Rev. C {\bf 87} (2012) 035206.

\bibitem{rocaoset} E.~Oset and L.~Roca, Phys. Rev. C {\bf 87} (2013) 055201.

\bibitem{prakhov} S.~Prakhov {\it et al.}  [Crystall Ball Collaboration],   Phys.\ Rev.\ C {\bf 70} (2004) 034605.

\bibitem{magas} V.~Magas, E.~Oset and A.~Ramos, Phys. Rev. Lett. {\bf 95} (2005) 052301.

\bibitem{oller2006} J.~A.~Oller, Eur. Phys. J. A {\bf 28} (2006) 63.

\bibitem{cosy} I.~Zychor {\it et al.}, Phys. Lett. B {\bf 660} (2008) 167.

\bibitem{hades} G.~Agakishiev {\it et al.} [HADES Collaboration] Phys. Rev. C {\bf 87} (2013) 025201.

\bibitem{geng}L.~S.~Geng and E.~Oset,  Eur.\ Phys.\ J.\ A {\bf 34} (2007) 405.

\bibitem{jido1}D.~Jido, E.~Oset and T.~Sekihara,  Eur.\ Phys.\ J.\ A {\bf 42} (2009) 257.

\bibitem{hyodo1} T.~Hyodo {\it et al.}, Phys. Rev. C {\bf 68} (2003) 065203.

\bibitem{braun}O.~Braun {\it et al.}, Nucl. Phys. B {\bf 129} (1977) 1.

\bibitem{thomas1}D.~W.~Thomas {\it et al.}, Nucl. Phys. B {\bf 56} (1973) 15.

\bibitem{curceanu} C.~Curceanu and J.~Zmeskal,  (eds.), {\it Mini-Proceedings of 
ECT Workshop ``Strangeness in Nuclei''}, 2011, arXiv:1104.1926.

\bibitem{prls}J.~A.~Oller, J.~Prades and M.~Verbeni,
  Phys.\ Rev.\ Lett.\  {\bf 96} (2006) 199202; Phys.\ Rev.\ Lett.\  {\bf 95} (2005) 172502.


\bibitem{ramos}E.~Oset and A.~Ramos,
  Nucl.\ Phys.\ A {\bf 635} (1998) 99.

\bibitem{nieves}C.~Garc\'{\i}a-Recio, M.~F.~M.~Lutz and J.~Nieves, Phys. Lett. B {\bf 582} (2004) 49; A. Ramos, 
E.~Oset and C.~Bennhold, Phys. Rev. Lett. {\bf 89} (2002) 252011.


\bibitem{morgan}K.~L.~Au, D.~Morgan and M.~R.~Pennington,
  Phys.\ Rev.\ D {\bf 35} (1987) 1633.



\bibitem{sekihara} T.~Sekihara, T.~Hyodo and D.~Jido, Phys. Lett. B {\bf 669} (2008) 133; Phys. Rev. 
C {\bf 83} (2011) 055202.

\bibitem{jidoroca}T.~Hyodo, D.~Jido and L.~Roca, Phys. Rev. D {\bf 77} (2008) 056010; 
L.~Roca, T.~Hyodo and D.~Jido, Nucl. Phys. A {\bf 809} (2008) 65.

\bibitem{goity}J.~L.~Goity, Phys. Atom. Nucl. {\bf 68} (2005) 624.


\end{thebibliography}
\end{document}